# Magnetic States and Electronic Properties of Manganese-Based Intermetallic Compounds Mn$_2$YAl and Mn$_3$Z
# ($Y$ = V, Cr, Fe, Co, Ni; $Z$ = Al, Ge, Sn, Si, Pt)

**Vyacheslav V. Marchenkov * and Valentin Yu. Irkhin**

M.N. Mikheev Institute of Metal Physics, UB RAS, 620108 Ekaterinburg, Russia;

*Correspondence: march@imp.uran.ru; valentin.irkhin@imp.uran.ru

**Abstract:** We present a brief review of experimental and theoretical papers on studies of electron transport and magnetic properties in manganese-based compounds Mn$_2$YZ and Mn$_3$Z ($Y$ = V, Cr, Fe, Co, Ni, etc.; $Z$ = Al, Ge, Sn, Si, Pt, etc.). It has been shown that in the electronic subsystem of Mn$_2$YZ compounds, the states of a half-metallic ferromagnet and a spin gapless semiconductor can arise with the realization of various magnetic states, such as a ferromagnet, a compensated ferrimagnet, and a frustrated antiferromagnet. Binary compounds of Mn$_3$Z have the properties of a half-metallic ferromagnet and a topological semimetal with a large anomalous Hall effect, spin Hall effect, spin Nernst effect, and thermal Hall effect. Their magnetic states are also very diverse: from a ferrimagnet and an antiferromagnet to a compensated ferrimagnet and a frustrated antiferromagnet, as well as an antiferromagnet with a kagome-type lattice. It has been demonstrated that the electronic and magnetic properties of such materials are very sensitive to external influences (temperature, magnetic field, external pressure), as well as the processing method (cast, rapidly quenched, nanostructured, etc.). Knowledge of the regularities in the behavior of the electronic and magnetic characteristics of Mn$_2$YAl and Mn$_3$Z compounds can be used for applications in micro- and nanoelectronics and spintronics.

**Keywords:** manganese based intermetallic compounds; antiferromagnetism; compensated ferrimagnetism; frustrated magnets; half-metallic ferromagnets; spin gapless semiconductors; topological semimetals; anomalous Hall effect; kagome lattice

## 1. Introduction

The exploration and advancement of novel materials with distinct magnetic and electronic properties, along with the experimental and theoretical investigation of their electronic energy spectrum, structural configurations, and magnetic states, hold significant importance from both a fundamental and a practical standpoint. Heusler compounds, discovered over 100 years ago by German chemist F. Heusler [1], continue to be a subject of great interest in current research. These compounds exhibit a wide range of unique multifunctional properties [2], such as half-metallic ferromagnetism [3–5], spin gapless semiconductor state [6–8], topological insulator and semimetal behavior [8–11], shape memory effect [12,13], magnetocaloric effect [14,15], and many others (see, for example, [2,16,17] and references therein).

Among these materials, intermetallic compounds based on manganese, such as Mn$_2$YZ and Mn$_3$Z ($Y$ = V, Cr, Fe, Co, Ni, etc.; $Z$ = Al, Ga, Ge, Sn, etc.), are particularly noteworthy. In addition to the functional properties mentioned above, these compounds exhibit a range of unique magnetic characteristics, including the ability to manifest states of antiferromagnetism, compensated ferrimagnetism, and frustrated magnets. They possess unconventional magnetic and electronic properties that are highly susceptible to external influences, making them promising candidates for practical applications in fields such as spintronics, microelectronics, and nanoelectronics.

Manganese-based Heusler compounds, specifically $Mn_2YZ$, possess several remarkable properties. They exhibit characteristics of half-metallic ferromagnets (HMF), spin gapless semiconductors (SGS), and potentially topological semimetals (TSM). These compounds also demonstrate a significant magnetocaloric effect and shape memory. Moreover, they offer the opportunity to realize unique magnetic states, including ferromagnetic (FM), antiferromagnetic (AFM), compensated ferrimagnetic (CFIM) ones, etc. (see, e.g., refs. [4,8,18]).

In previous studies, the intermetallic compounds $Mn_3Z$ ($Z$ = Ge, Sn, Ga, Ir, Rh) were found to exhibit a strong anisotropic anomalous Hall effect (AHE) and spin Hall effect in their antiferromagnetic state [19]. Researchers reported [20] a zero magnetic moment in thin films of the $Mn_3Al$ alloy, which was attributed to a compensated ferrimagnetic state. This CFIM state differs from AFM due to the distinct crystallographic positions of manganese. Another study [21] observed a zero magnetic moment in the cast $Mn_3Al$ alloy and suggested that it may also be indicative of compensated ferrimagnetism.

The combination of noncollinear magnetic structure and Berry curvature can result in a nonzero topological anomalous Hall effect, as demonstrated in antiferromagnets $Mn_3Sn$ and $Mn_3Ge$ [9,22]. These compounds, along with their noncollinear magnetic structures, also exhibit topological states in real space in the form of magnetic anti-skyrmions. The ability to manipulate the Berry curvature highlights the significance of understanding both the electronic and magnetic structures of $Mn_3Z$ compounds.

Furthermore, among these intermetallic compounds, there are topological systems that possess unique surface states and display anomalous transport phenomena due to their unconventional bulk electronic topology. For instance, $Mn_3Ge$ and $Mn_3Sn$ compounds with a distorted $D0_{19}$ structure in the ab plane form a lattice of Mn atoms resembling a highly frustrated kagome lattice [22,23]. Further theoretical and experimental investigations of such structures and their impact on electronic and magnetic properties remain pertinent and in demand.

Figures 1–3 schematically show the models of an antiferromagnet and a compensated ferrimagnet (Figure 1), a half-metallic ferromagnet and a spin gapless semiconductor (Figure 2), and topological materials (Figure 3).

The magnetic structure of $D0_3$ compounds in the antiferromagnetic and compensated ferrimagnetic states is presented in Figure 1. As can be seen from Figure 1, the difference between antiferromagnetic and ferrimagnetic states is that the crystallographic positions of manganese with oppositely directed magnetic moments are completely different.

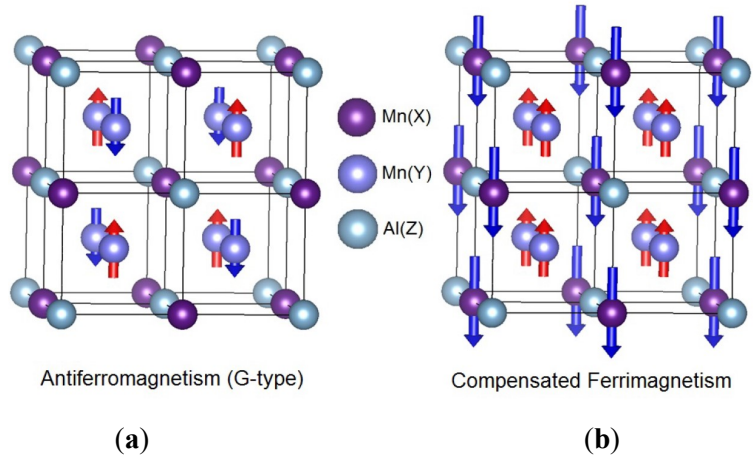

(a)                    (b)

**Figure 1.** Schematic view of the magnetic structure of $D0_3$ compounds. (**a**) G-type antiferromagnetic structure of $V_3Al$, ions in Mn($X$) positions being nonmagnetic. (**b**) Magnetic structure of compensated ferrimagnet $Mn_3Al$ [20]. Up-directed magnetic moments are shown by red arrows, and down-directed ones by blue.

Figure 2 shows the states of a half-metallic ferromagnet and a spin gapless semiconductor. The electronic structure of a half-metallic ferromagnet has the following features: there is a gap at the Fermi level for spin-down electronic states, which is absent for spin-up ones (Figure 2a). In the case of a spin gapless semiconductor (Figure 2b), the situation is similar, but there is a significant difference. As in a half-metallic ferromagnet, there is a finite gap for the spin-down spin projection, but the gap is zero for spin-up ones (Figure 2b). This is similar to the case of classical gapless semiconductors [24].

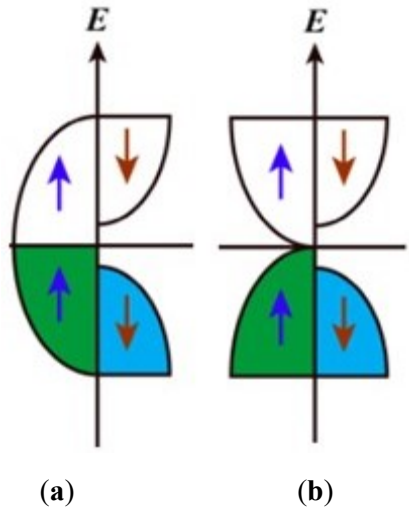

(a)           (b)

**Figure 2.** Simple models of half-metallic ferromagnets (**a**) and spin gapless semiconductors (**b**) [8]. The valence band for the electron states with spin up (blue arrows) is marked by green, and for ones with spin down (red arrows) by blue.

In topological materials (Figure 3), an inversion of the conduction and valence bands can occur because of a strong spin–orbit interaction. This leads to the appearance of a nontrivial topology of the electronic band structure, which is observed in topological insulators, Dirac and Weyl semimetals. Topological insulators have a characteristic energy gap in bulk and "metallic" states on the surface (Figure 3a). The Dirac and Weyl semimetals also have a gap in the bulk resulting from strong spin–orbit coupling, except for the band intersection at Dirac (Figure 3b) and Weyl (Figure 3c) points, respectively.

The methods used to prepare intermetallic compounds, specifically those dealing with $Mn_3Z$, can have a significant impact on their structural properties (see, e.g., refs. [25–29]). This, in turn, influences their electronic and magnetic states. Understanding the role of the structural state in the formation and behavior of intermetallic compounds based on manganese is an important and fascinating problem.

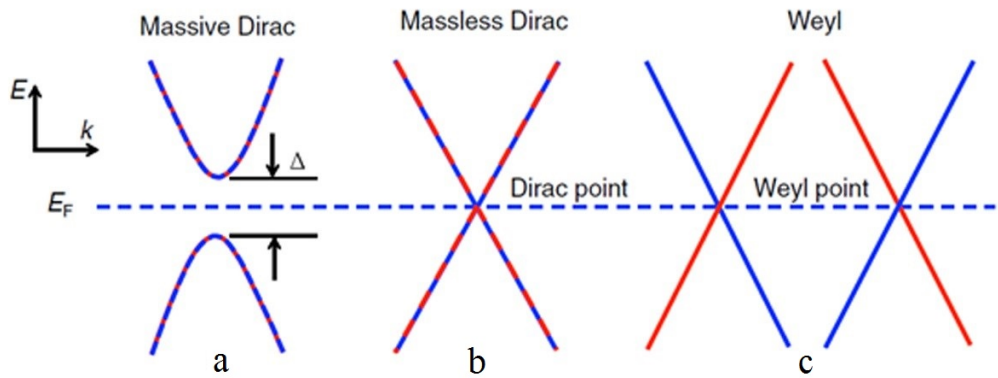

**Figure 3.** Schematic representation of the band structure of (**a**) massive Dirac, (**b**) "massless" Dirac, and (**c**) Weyl fermions. The latter arises during the decay of the Dirac point. Two-color and one-color curves and lines represent doubly degenerate and nondegenerate zones, respectively [8].

Both experimental and theoretical studies in this area hold great relevance and scientific significance (see papers [30–32] and references therein). They allow for a comprehensive description of the evolution of the electronic structure, magnetic state, and properties of $Mn_2YZ$ and $Mn_3Z$ compounds. These studies aim to understand the characteristics and differences in the manifestation of various states, such as antiferromagnetism, compensated ferrimagnetism, frustrated magnetism, half-metallic ferromagnetism, spin gapless semiconductors, topological semimetals, etc. This work provides a concise overview of the research on electron transport and the magnetic state in compounds based on manganese, specifically $Mn_2YZ$ and $Mn_3Z$ ($Y$ = Sc, Ti, V, Cr, Fe, Co, etc.; $Z$ = Al, Ge, Sn, Si, Pt, etc.).

**2. Electronic and Magnetic Properties of $Mn_2YZ$ ($Y$ = V, Cr, Fe, Co, Ni, etc.; $Z$ = Al, Ge, Sn, Si, Pt, etc.) Compounds**

The HMF state was theoretically predicted in 1983 in ref. [33] using band calculations of the NiMnSb half-Heusler compound. Later, the HMF theory was significantly supplemented and extended [3,34]. After 40 years, band calculations are still extensively used to predict the HMF state in various compounds (see, for example, refs. [4,5] and references therein).

Electronic structure calculations performed on manganese-based Heusler alloys have shown that many of them can exhibit HMF properties. These are $Mn_2VZ$ ($Z$ = Al, Ga, In, Si, Ge, Sn) compounds [35], $Mn_2FeZ$ ($Z$ = Al, Ga, Si, Ge, Sb) [36], $Mn_2CrZ$ ($Z$ = Al, Ga, Si, Ge, Sb) [37], $Mn_2TiZ$ ($Z$ = Al, As, Bi, Ga, Ge, Sb, Si, Sn) [38], $Mn_2ZrZ$ ($Z$ = Ga, Ge, Si) [39,40]. It is worth emphasizing that the HMF state predicted theoretically and/or as a result of band calculations is by no means always realized in real compounds.

Using density functional theory, the electronic, magnetic, and structural properties of ferrimagnetic $Mn_2VAl$ and $Mn_2VSi$ Heusler alloys were studied in ref. [41]. It was shown that two states can exist in studied compounds: one state with low magnetization and a HMF behavior with almost 100% spin polarization and the second state with high magnetization and a metallic character.

The temperature dependences of the electrical resistivity $\rho(T)$ of the $Mn_2CrAl$ alloy were studied in [21]. It can be observed (see Figure 1 in ref. [21]) that the residual resistivity $\rho_0$ reaches a large value of ≈250 $\mu\Omega$ cm, and the resistivity $\rho$ decreases with temperature increasing up to room temperature. A similar behavior $\rho(T)$ was observed for $Mn_2FeAl$ in [42]. The large values of $\rho_0$ and the "semiconductor" dependence $\rho(T)$ were explained by the structural disorder. There is a so-called Mooij rule [43], according to which in metallic systems with the static disorder, i.e., with resistivity $\rho$ > (150–200) $\mu\Omega$ cm, a negative temperature coefficient of resistance is usually observed. The estimates of the concentration $n$ of current carriers from measurements of the Hall effect give large values: $n \approx 1.6 \cdot 10^{22}$ cm$^{-3}$ for $Mn_2CrAl$ [21], and $n \approx 2 \times 10^{22}$ cm$^{-3}$ for $Mn_2FeAl$ [42].

Studies of the magnetic properties of $Mn_2CrAl$ [21] and $Mn_2FeAl$ [42,44] alloys showed that the behavior of the magnetization $M(H)$ indicates a total moment close to zero (see, e.g., Figure 4). Such a state can be characterized as (1) a compensated ferrimagnet, which retains the nature of a half-metallic ferromagnet state [20,45], or as (2) an antiferromagnet.

The distinction between them can be explained as follows. In the case of antiferromagnetism, the opposite magnetic moments of manganese are situated in positions that are equivalent from a crystallographic perspective. Conversely, in the case of compensated ferrimagnetism, these opposite moments of manganese occupy crystallographically distinct positions. In CFIM, the compensation of magnetic moments is not solely reliant on crystal symmetry but is also influenced by a unique band structure. The utilization of CFIM and/or AFM states in such materials holds promise for applications in spintronics as they can exhibit high spin polarization of current carriers.

It is shown in ref. [44] that a frustrated antiferromagnetic state is observed in $Mn_2FeAl$ alloy with a Neel temperature $T_N$ = 48 K and a Curie–Weiss temperature $\theta_{CW} \approx -230$ K. In this case, large antiferromagnetic spin fluctuations, caused by the geometric frustrations, lead to the appearance of an unusually large electronic heat capacity. According to ref. [44], the corresponding value of the $T$-linear heat capacity coefficient is $\gamma$ = 210 mJ/molK$^2$.

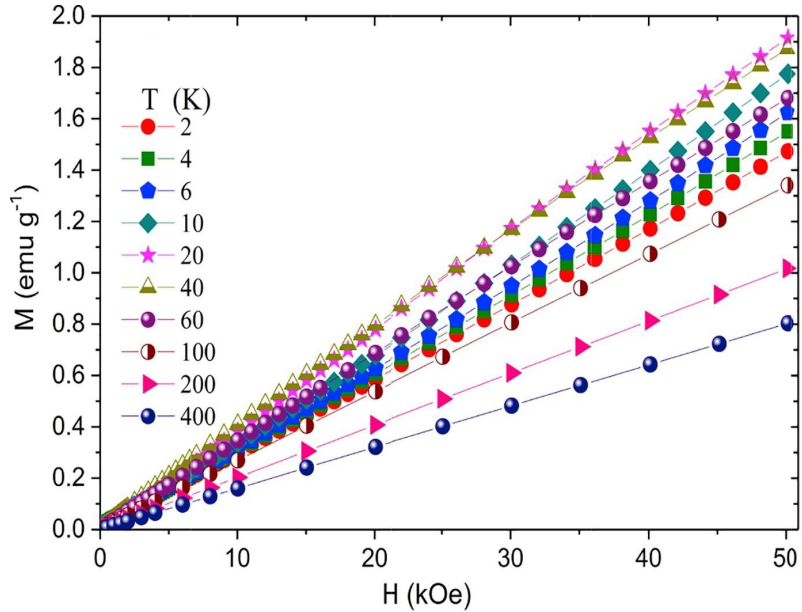

**Figure 4.** Field dependence of magnetization for $Mn_2FeAl$ at different temperatures [42].

The $Mn_2CoAl$ compound exhibits even higher residual resistivity values [46,47] (Figure 5), belonging to a novel class of quantum materials known as spin gapless semiconductors. The theoretical prediction of SGS materials was made in 2008 [6]. SGS materials possess an unusual band structure, wherein there exists an energy gap for the spin-down electron subsystem near the Fermi level, while the spin-up electronic states have the top of the valence band touching the bottom of the conduction band. Under these circumstances, the kinetic properties of SGS materials are expected to resemble those of "classical" gapless semiconductors [24], including high residual resistivity with weak temperature dependence, relatively low current carrier concentration, and low thermoelectric power. Due to the prevalence of spin-up current carriers, a high degree of spin polarization, a significant magnetization, and an anomalous Hall effect are anticipated. Experimental evidence of this spin gapless semiconductor state with unusual magnetic and magnetotransport properties has been observed in $Mn_2CoAl$ [46,47] and $Ti_2MnAl$ Heusler alloys [48].

Figure 5 shows the temperature dependencies of the electrical resistivity, the Seebeck coefficient, and the concentration of current carriers of the $Mn_2CoAl$ compound [47]. It can be seen that the residual resistivity is very large and amounts to ~445 μΩ cm, and the resistivity itself slightly decreases with temperature, reaching a value of ~410 μΩ cm at room temperature.

In this case, the thermoelectric power is small and weakly changes with temperature. Finally, the concentration of charge carriers is relatively low, monotonically increasing with temperature: $n \approx 1.65 \times 10^{20}$ cm$^{-3}$ at 2 K and $\sim 3 \times 10^{20}$ cm$^{-3}$ at room temperature. Finally, measurements of the Hall effect in Mn$_2$CoAl [47] show that the value of the anomalous Hall conductivity $\sigma_{xy}$ is relatively small and at 2 K is $\sigma_{xy} = 21.8$ S cm$^{-1}$, which can be explained by the symmetry features of the Berry curvature (Figure 5). These facts confirm the realization of the state of a spin gapless semiconductor in the Mn$_2$CoAl compound.

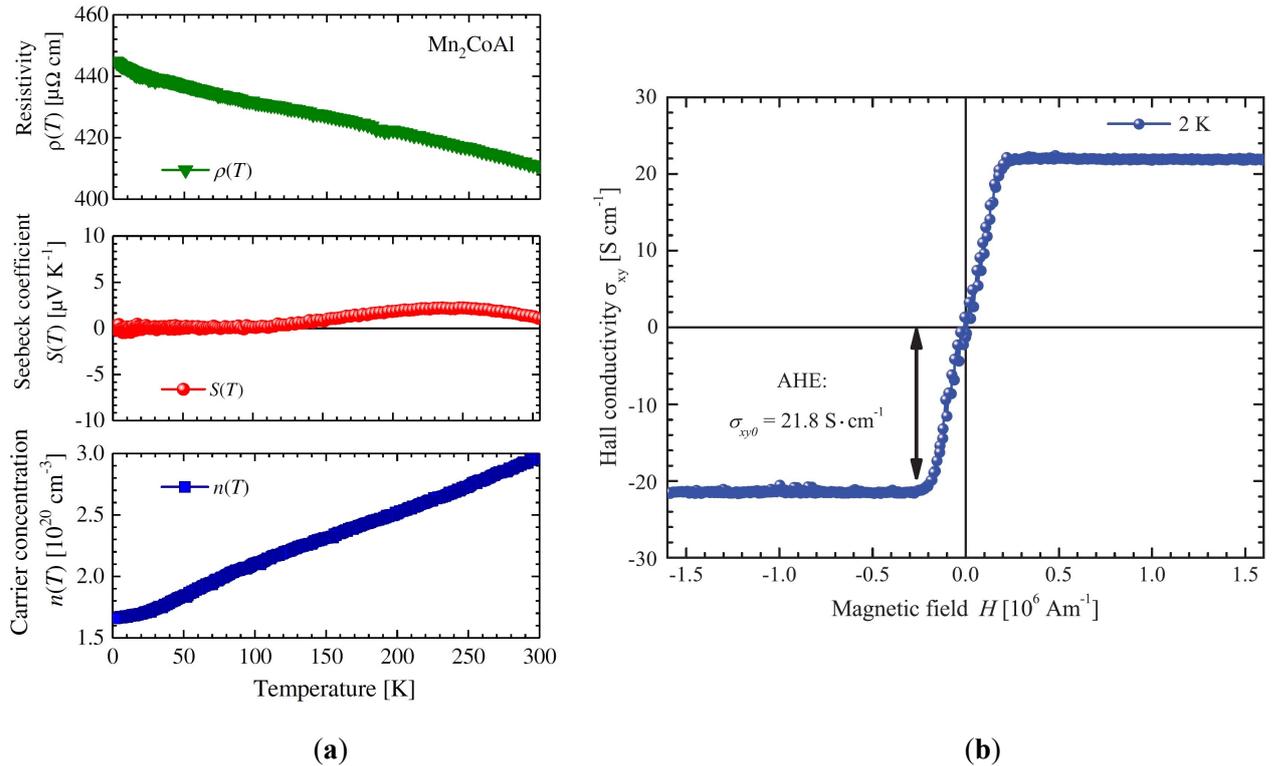

(a)                  (b)

**Figure 5.** Mn$_2$CoAl compound [47]. (**a**) Temperature dependence of the electroresistivity $\rho$, Seebeck coefficient $S$, and charge carriers concentration $n$. (**b**) Field dependence of Hall conductivity at $T = 2$ K.

It is important to highlight that the characteristics of a spin gapless semiconductor (SGS) are not limited to electron transport but also extend to other properties, including optical characteristics. In a study referenced as [49], the optical properties of the Mn$_2$CoAl compound, which exhibits SGS behavior, were investigated. The study also involved the calculation of its electronic structure. The findings revealed distinct optical properties of Mn$_2$CoAl, including positive values of the real part of the dielectric constant ($\varepsilon_1$) and the absence of the Drude contribution to the optical conductivity in the infrared (IR) region of the spectrum within the examined range. These observations suggest a slight deviation or "deterioration" of the metallic properties of Mn$_2$CoAl.

Besides, the authors of the study observed [49] intense interband absorption in the IR range and concluded that the band spectrum had a complex structure with a high density of $d$-states near the Fermi level $E_F$. The observed features of the optical properties allowed the authors [49] to explain the band spectrum characteristic of spin gapless semiconductors.

The electronic structure and magnetic properties of Heusler alloys, specifically Mn$_2YZ$, can be significantly altered by modifying their composition through the substitution of atoms and applying external influences, like hydrostatic pressure and mechanical processing. In ref. [50], the authors investigated the structure, mechanical, electronic, and magnetic properties of Mn$_{2-x}$Fe$_{1+x}$Si Heusler alloys (where $x$ = 0, 0.25, 0.5, 0.75, 1) using density functional theory. By replacing Mn atoms with Fe atoms, it was observed that all the studied alloys maintain mechanical and dynamic stability, adopting the Hg$_2$CuTi structure for $x$ = 0, 0.25 and 0.5 and the Cu$_2$MnAl structure for $x$ = 0.75 and 1. Furthermore, as the iron content increases, the alloys exhibit improved ductility. All the studied alloys display half-metallic ferromagnetic behavior, with the Fermi levels gradually shifting toward the middle of the bandgap in the spin-down direction with increasing iron content.

In the study [51], researchers successfully synthesized the Mn$_2$FeSi Heusler alloy using a high-energy planetary ball mill. The resulting compound was found to be an inverse Heusler alloy with an X$_A$ structure. Through the use of Mössbauer spectroscopy and magnetization measurements, they discovered that the synthesized material exhibits a heterogeneous magnetic structure at room temperature. This structure is composed primarily of the paramagnetic phase, with small contributions from ferro- and ferrimagnetic phases. The Neel temperature $T_N$ = 67 K was determined from the temperature dependences of the magnetization.

Using ab initio calculations and the Monte Carlo simulation method in [52], the structural, magnetic, and electronic properties of Mn$_2Y$Sn ($Y$ = Sc, Ti, and V) Heusler alloys were studied under applied hydrostatic pressure. The coexistence of two magnetic states with a low and a high magnetic moment was found for a small and large volume of the crystal lattice, correspondingly. These states coexist together due to almost equal energy at an applied pressure of 3.4, −2.9, and −3.25 GPa for Mn$_2$ScSn, Mn$_2$TiSn, and Mn$_2$VSn, respectively. A positive pressure corresponds to a uniform compression of the lattice, and a negative one corresponds to a uniform expansion of the lattice. It was demonstrated that for the studied compounds, the low-magnetic state (LMS) was characterized by an almost half-metallic behavior, while the high-magnetic state (HMS) acquired a metallic character. It was shown that the parameters of magnetic exchange and Curie temperatures were significantly higher for the HMS than for the LMS. The authors proposed a mechanism for switching between the half-metallic state with the LMS and the metallic state with the HMS using applied pressure.

In addition to the electronic and magnetic states discussed above, Heusler alloys based on manganese can exhibit other thermodynamic phenomena, such as the shape memory effect and the magnetocaloric effect. Moreover, unlike the well-known Heusler-like alloys based on Ni-Mn-$Z$ ($Z$ = Ga, In, Sn, etc.) with strong deviations from stoichiometry (of the type Ni$_{2\pm x}$Mn$_{1\pm x}$$Z$), SME and MCE can also be observed in full Heusler alloys Mn$_2$$YZ$, where $Y$ = Sc, Ti, V, Co, Ni, etc.; $Z$ = Ga, Ge, Sn, etc. (see, e.g., [53–55] and references therein).

Similar magnetic and electronic states can be realized in binary intermetallic compounds based on manganese Mn$_3$$Z$ ($Z$ = Al, Ge, Ga, etc.) as well.

### 3. Features of the Electronic Transport and Magnetic State of Mn$_3$$Z$ ($Z$ = Al, Ge, Si, Sn, Pt, etc.) Compounds

Compensated ferrimagnets possess a magnetic moment that totals zero, but their density of electronic states for opposite spin directions allows for a significant spin polarization of current carriers. A study [20] indicates that the compensated ferrimagnetic state is present in Mn$_3$Al, and this half-metallic state can persist at room temperature. In the same study, the magnetic properties of thin films of Mn$_3$Al were examined. The results reveal that the films demonstrate the state of a compensated ferrimagnet, with a total magnetization value of $M$ = 0.11 ± 0.04 μ$_B$/f.u.

Results [20] are in good agreement with refs. [28,29] where the electronic transport and magnetic characteristics of cast and rapid melt-quenched (RMQ) Mn$_3$Al alloys were measured and compared.

In Figure 6, the field dependence of the magnetization ($M$ = f($H$)) is shown for both the cast and RMQ alloy Mn$_3$Al at a temperature of 4.2 K, as described in reference [29]. It can be observed from Figure 6a that the cast alloy exhibits a small magnitude of magnetization, which increases linearly with the applied field. At a field strength of 70 kOe, the magnetization reaches approximately 1.1 emu/g. In contrast, the Mn$_3$Al RMQ alloy (Figure 6b) displays a significantly different shape of the $M(H)$ dependence. In this case, even in weak magnetic fields, there is an increase in magnetization, followed by a tendency toward saturation.

In ref. [29], the difference in $M(H)$ behavior for Mn$_3$Al is explained as follows. In the case of a cast compound having the $β$-Mn structure, a frustrated antiferromagnetic state can arise, which manifests itself in the $M(H)$ dependences (Figure 6a) and in the temperature dependences of the susceptibility (Figure 7). The results obtained in [29] are in good agreement with the conclusions of ref. [42] obtained on Mn$_2$FeAl with the same $β$-Mn structure.

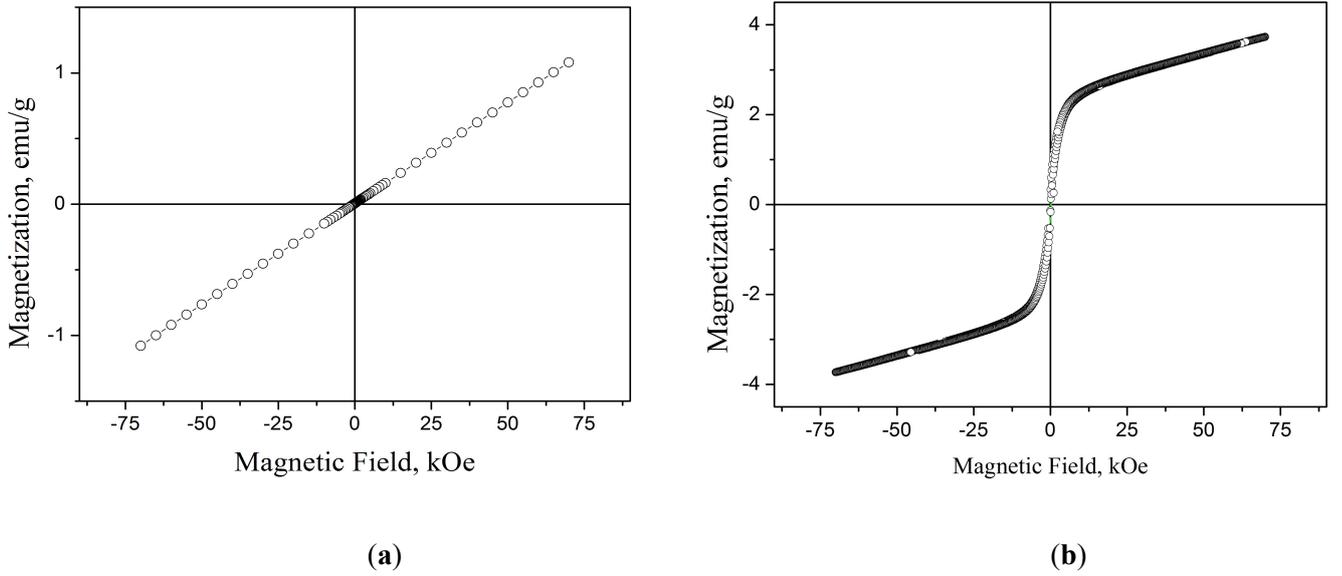

**Figure 6.** Field dependence of magnetization for Mn$_3$Al at $T$ = 4.2 K: (**a**) cast and (**b**) RMQ samples [29].

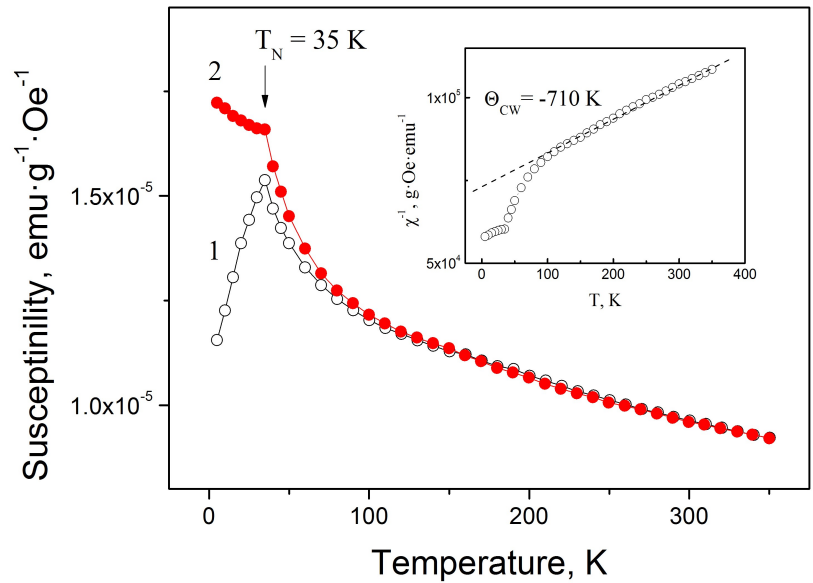

**Figure 7.** The magnetic susceptibility of the Mn$_3$Al alloy measured at a magnetic field strength of 100 Oe during the cooling process. The results are represented by two sets of data points: (1) open circles that indicate measurements conducted without the application of a magnetic field and (2) solid red circles that correspond to the presence of a magnetic field. The inset shows the temperature dependence of the inverse magnetic susceptibility $\chi^{-1}$ [29]. The Curie–Weiss law with the temperature $\theta_{CW} \approx -710$ K is shown by dashed line.

According to ref. [29], a decrease in grain size and an increase in "disorder" are observed in Mn$_3$Al after RMQ treatment. As shown in [56], the magnetic state of the Mn$_3$Al alloy is very sensitive to the filling of interstices and degree of ordering. According to first-principles calculations [18] in Mn$_3$Al with the $D0_3$ structure, a compensated ferrimagnetic state with a zero moment can arise, which is observed in the RMQ compound Mn$_3$Al (Figure 6b) according to [29].

Figure 7 displays the temperature dependence of the susceptibility for the Mn$_3$Al alloy [29]. The inset graph illustrates the temperature dependence of the inverse susceptibility at a magnetic field strength of 100 Oe. At high temperatures, the data approximately follow the Curie–Weiss law with a temperature θ$_{CW}$ of around −700 K. This behavior suggests the possible presence of an antiferromagnetic state. The Neel temperature $T_N$ can be estimated by observing the point where the magnetic susceptibility curve breaks. In our case, $T_N$ is determined to be 35 K (as shown in Figure 7). The large ratio |θ$_{CW}$|/$T_N$ indicates a frustrated antiferromagnetism in the cast Mn$_3$Al compound [29].

The cast and RMQ alloys investigated in [28,29] have the *β*-Mn structure. This consists of two distinct sublattices, one of which is made up of triangles arranged perpendicular to the [111] axes and forming a three-dimensional kagome-like lattice. Recent experimental studies have shown that in systems with strong frustration caused by competing exchange interactions, not only a quantum state of spin liquid can arise, but also antiferromagnetism with a significantly reduced, but still finite, Neel temperature can occur. These systems are characterized by a frustration parameter, the ratio |θ$_{CW}$|/$T_N$; in the intermediate temperature range $T_N < T < $|θ$_{CW}$|, the system can exhibit unusual spin-liquid properties. High values of the frustration parameter have been observed in the PdCrO$_2$ compound where $T_N$ = 37 K and θ$_{CW}$ ≈ −500 K [57]. This behavior is not explained by the standard Heisenberg model and is thought to be due to correlation effects in the subsystem of itinerant electrons [58]. A similar behavior of the magnetic susceptibility was recently discovered in the Mn$_2$FeAl compound with the *β*-Mn structure in [42,44] ($T_N$ = 42 K, θ$_{CW}$ ≈ −230 K according to [44]).

According to calculations [56], the Mn$_3$Al compound in the *β*-Mn structure exhibits a ferrimagnetic state, where the magnetic moment of the sublattices is not significantly compensated. However, ab initio calculations [20,59] in the $D0_3$ structure suggest that a compensated ferrimagnetic state with a zero moment and a half-metallic structure arises. The results from [60] are not fully supported by experimental data [28,29]. These data suggest that both the cast and RMQ Mn$_3$Al alloys exhibit a frustrated antiferromagnetic state and an almost compensated ferrimagnetic state, respectively. In the case of a cast Mn$_3$Al alloy with the *β*-Mn structure, the absence of a long-range magnetic order can manifest itself as the Hall effect in the form of a zero anomalous contribution. On the contrary, in the case of the RMQ alloy with the *β*-Mn structure in the state of a compensated ferrimagnet, an anomalous Hall effect should be observed.

Figure 8 illustrates the temperature dependence of the electrical resistivity $\rho(T)$ for cast and rapidly melt-quenched Mn$_3$Al alloys [29]. In the cast alloy (Figure 8a), it can be observed that the residual resistivity $\rho_0$ is relatively high, reaching a value of approximately 307 μΩ cm. The resistivity follows a semiconductor-like trend, decreasing with increasing temperature. However, after quench hardening (Figure 8b), there is a significant reduction in the residual electrical resistivity $\rho_0$ by more than an order of magnitude, down to 12 μΩ cm. Additionally, a minimum point appears on the temperature dependence $\rho(T)$ at around 60 K.

Typically, rapid melt quenching results in the formation of fine-grained structures, which can lead to increased resistivity due to the scattering of current carriers at grain boundaries. However, this phenomenon was not observed in the Mn$_3$Al alloy studied [28,29]. Instead, the presence of manganese sulfide (MnS) precipitates at the grain boundaries occurred, explaining the high electrical resistivity values in the cast alloy. The rapid melt-quenching process caused the dissolution of manganese sulfide within the volume of the grains, resulting in grain boundaries free of MnS. As a result, the resistivity of the rapidly melt-quenched alloy decreased compared with the cast alloy. Similar behavior was also observed in the Hall effect measurements [28,29].

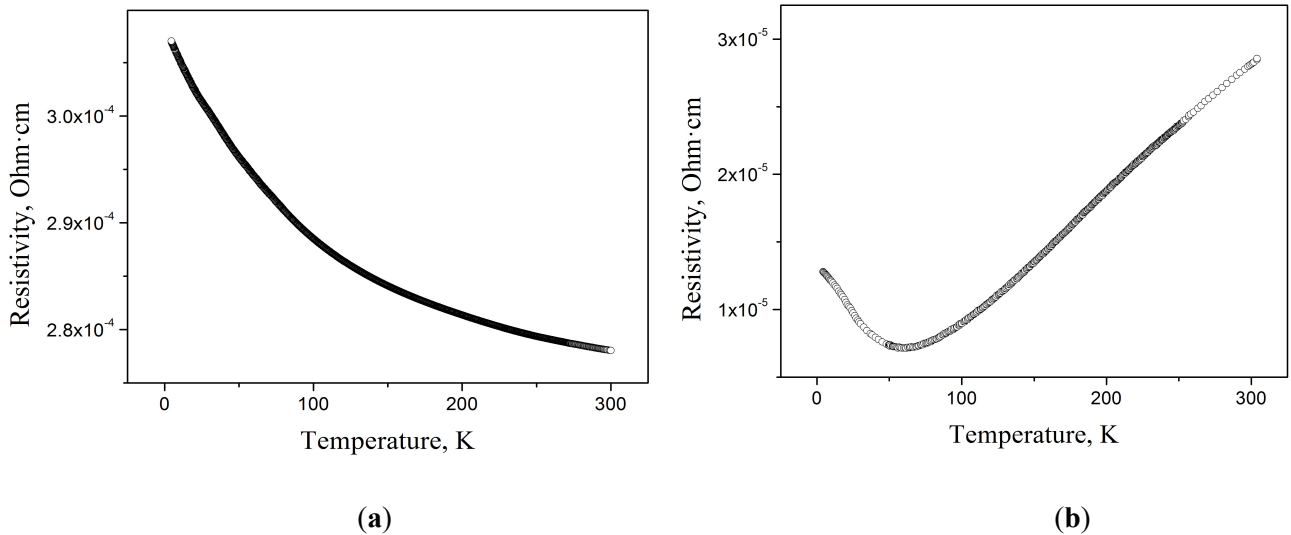

(a)            (b)

**Figure 8.** Temperature dependence of the resistivity for cast (**a**) and RMQ (**b**) Mn$_3$Al [29].

Figure 9 shows the dependence of the Hall resistivity $\rho_{xy}$ = f($H$) at $T$ = 4.2 K for the cast and RMQ Mn$_3$Al alloy. In the first case, a linear increase in $\rho_{xy}$ is observed, and in the second case, the behavior of $\rho_{xy}(H)$ is typical for alloys with an anomalous Hall effect [28,29]. These data also indicate the absence of spontaneous magnetization in the case of cast Mn$_3$Al and the realization of a compensated ferrimagnet state in the RMQ Mn$_3$Al alloy [28,29].

Traditionally, it was believed that the anomalous Hall effect (AHE) was exclusive to ferromagnetic materials [61]. However, it has been discovered that the AHE can manifest in various magnetic materials as a result of broken time-reversal symmetry. In reference [62], ab initio calculations were conducted, taking into consideration the symmetry properties of the nontrivial AHE in the $Mn_3Al$ compensated ferrimagnet. The nonzero elements of the anomalous Hall conductivity were determined based on the magnetic space group of $Mn_3Al$. The calculations revealed that the value of the anomalous Hall conductivity was $\sigma_{xy} = -320\ (\Omega\ cm)^{-1}$. The study also delved into the nature of the Berry curvature, which is responsible for the internal origin of the AHE, using group theory. Furthermore, the overall behavior of the Berry curvature across the entire Brillouin zone was illustrated.

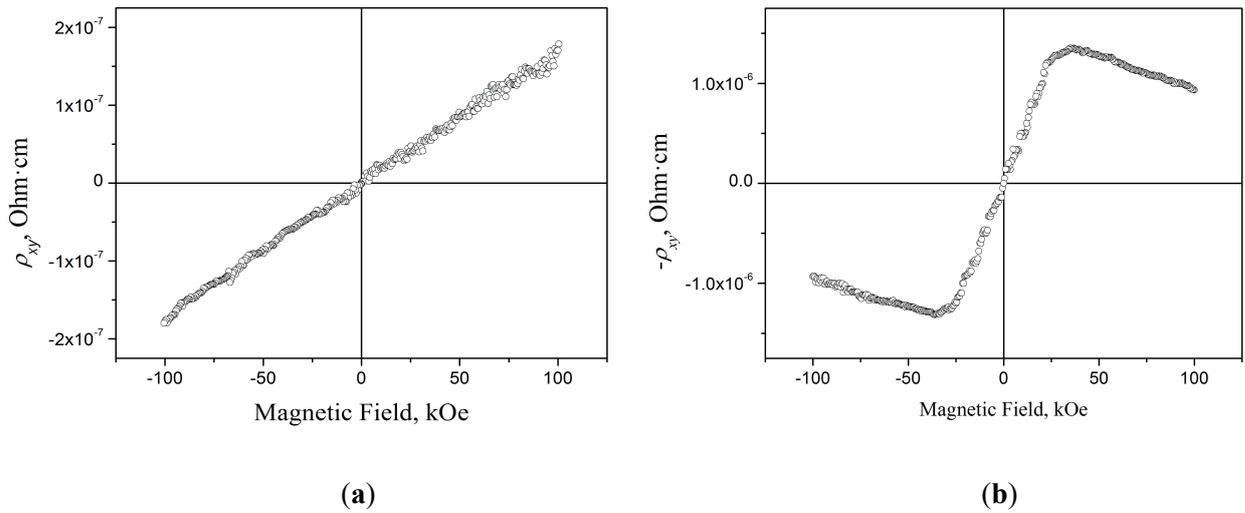

(a) (b)

**Figure 9.** Field dependence of Hall resistivity for $Mn_3Al$ at $T = 4.2$ K [29]: (**a**) cast and (**b**) RMQ samples.

The observation of a nonzero topological anomalous Hall effect in antiferromagnets $Mn_3Sn$ and $Mn_3Ge$ [9] can be attributed to their noncollinear magnetic structure and Berry curvature. A study conducted in [63] focused on the magnetization and anomalous Hall effect of a single-crystal hexagonal chiral antiferromagnet $Mn_3Ge$. Remarkably, it was discovered that this material exhibited a significant anomalous Hall conductivity of approximately $60\ (\Omega\ cm)^{-1}$ at room temperature and $\sim 380\ (\Omega\ cm)^{-1}$ at 5 K in zero field, approaching half the anticipated value for a quantum Hall effect on an atomic layer with a Chern number of one. Furthermore, the sign of the anomalous Hall effect was found to change with a magnetic field vector direction change of less than 0.1 T or with a rotation of a small magnetic field of less than 0.1 T. This intriguing behavior could have implications in the development of switching and storage devices based on antiferromagnets.

It is worth noting that recent studies have reported new antiferromagnetic materials with rapid response times, low power consumption, and high immunity to external magnetic fields. These materials show promise in the field of spintronics [64–69].

In materials exhibiting noncollinear antiferromagnetism, the magnetic moments of atoms are not strictly aligned along a single axis. One such material is Mn$_3$Ge, which falls into the category of noncollinear antiferromagnetic materials. In Mn$_3$Ge, the manganese atoms form a hexagonal sublattice, and their magnetic moments arrange themselves in a kagome lattice structure (Figure 10). This unique arrangement gives rise to a noncollinear antiferromagnetic state and several other unusual effects, including a large anomalous Hall effect, a new spin Hall effect, and a significant spin Nernst effect [70–72].

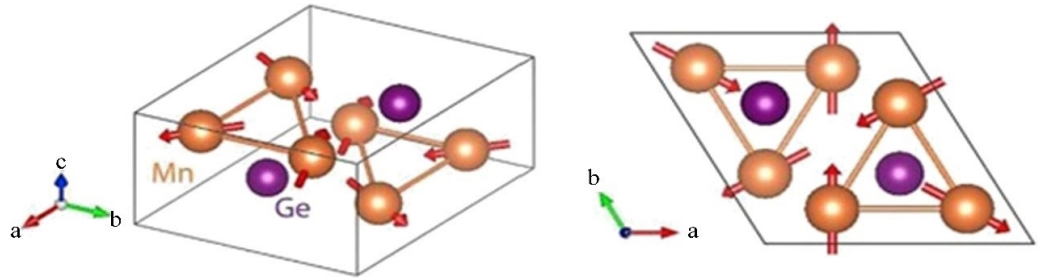

**Figure 10.** $D0_{19}$ structure for Mn$_3$Ge. Red arrows show the Mn magnetic moments [23].

A study conducted in [73] focused on the magnetic structure of antiferromagnetic Mn$_3$Ge crystals with a kagome lattice, using polarized neutron diffraction. The study has revealed that in Mn$_3$Ge, the magnetic order is characterized by a coplanar state represented by **k** = 0, which belongs to the two-dimensional irreducible representation of the double group $\Gamma_9$, which is the only irreducible representation consistent with the observed diffraction pattern of Mn$_3$Ge [73]. This coplanar state exhibits a perfect 120° antichiral structure with an angle of 120° and a magnetic moment of 2.2 μ$_B$/Mn. Additionally, a weak collinear ferromagnetism is observed. A phenomenological spin Hamiltonian is proposed to describe the manganese-based magnetism, incorporating exchange interactions, Dzyaloshinskii–Moriya interactions, and single-ion crystal field terms. These interactions contribute to spin wave damping and an extended range of magnetic interactions indicate an itinerant magnetism compatible with anomalies of transport properties.

In [74], a study was conducted on the electrical, magnetic, galvanomagnetic, and thermal properties of a single crystal of the noncollinear antiferromagnet Mn$_3$Ge. The researchers discovered that at very low temperatures, the Wiedemann–Franz law, which establishes a connection between the electronic and thermal Hall effects, holds true. However, deviations from this law were observed at temperatures above 100 K. The investigation revealed that the carriers in Mn$_3$Ge have a short mean free path, similar to the distance between Mn antisite defects. As a result, the elastic scattering of current carriers becomes the dominant factor. The researchers proposed that the deviation from the Wiedemann–Franz law is attributed to a disparity between the thermal and electrical contributions of the Berry curvature along the Fermi surface. Theoretical calculations supported this interpretation, demonstrating that the Berry spectra in the two systems are not identical. Importantly, the study also confirmed that the Bridgman relation, which links the anomalous Nernst and Ettingshausen coefficients, is valid across the entire temperature range investigated.

External pressure and doping can alter the Berry curvature, as well as the corresponding physical properties. In a study [75], the anomalous Hall effect in a single crystal of antiferromagnetic $Mn_3Ge$ was investigated under hydrostatic pressure up to 2.85 GPa. It was observed that the Hall signal becomes zero at a pressure of 1.53 GPa, and changes sign at higher pressures. Although the sign of the Hall conductivity changes with increasing pressure, its saturation value at room temperature remains relatively high, around 23 $(\Omega\ cm)^{-1}$ at 2.85 GPa, which is comparable to the saturation value at atmospheric pressure 40 $(\Omega\ cm)^{-1}$. The authors suggest that this change in the Hall conductivity may be attributed to gradual modifications in the in-plane components of the Mn moments within a noncollinear triangular magnetic lattice. These findings provide insights into the possibility of manipulating and controlling the anomalous Hall effect in chiral antiferromagnetic $Mn_3Ge$ through pressure-induced changes.

Authors of [76] investigated the magnetic and electronic structure of the $(Mn_{0.78}Fe_{0.22})_3Ge$ hexagonal single crystal, which is known as a Weyl semimetal, using electron transport, magnetic properties, and neutron diffraction experiments. The researchers conducted temperature measurements of the magnetization and observed two magnetic transitions at $T_{N1}$ = 242 K and $T_{N2}$ = 120 K. In this case, the anomalous Hall effect is observed in the intermediate range at temperatures between $T_{N1}$ and $T_{N2}$, disappearing at $T < T_{N2}$. Neutron diffraction experiments were carried out to determine the magnetic structures of the $(Mn_{0.78}Fe_{0.22})_3Ge$ single crystal. Neutron diffraction studies made it possible to conclude that the sample has a collinear antiferromagnetic structure below the temperature $T_{N2}$. In this case, the magnetic structure is noncollinear antiferromagnetic in the intermediate temperature range. The authors conclude that the observation of an anomalous Hall effect and a noncollinear magnetic structure in $(Mn_{0.78}Fe_{0.22})_3Ge$ in this temperature range indicates the existence of Weyl points. At temperatures below $T_{N2}$, there is no anomalous Hall effect, and the magnetic structure changes to a collinear antiferromagnetic one. This indicates a strong coupling between the magnetic and electronic structures of the $(Mn_{0.78}Fe_{0.22})_3Ge$ compound.

The influence of pressure, reaching up to 2.2 GPa, on the electrical resistivity $\rho$ and thermoelectric power $S$ of the $Mn_3Si$ single-crystal compound was investigated in a study referenced as [77]. The Neel temperature $T_N$ was determined by analyzing the temperature dependencies of $\rho$ and $S$, revealing an increase in $T_N$ with rising pressure. Notably, the resistivity and thermoelectric power displayed significant changes at around $P \approx 1$ GPa at a temperature of 2 K, indicating the occurrence of a phase transition.

Another research, referenced as [78], presents findings on the structural and magnetic properties of noncollinear antiferromagnetic $Mn_3Sn$ films that adopt the $D0_{19}$ hexagonal structure. These films were observed to exhibit weak ferromagnetism, characterized by an uncompensated in-plane magnetization of 34 kA/m and a coercive force $\mu_0H_c$ of 4.0 mT at room temperature. Additionally, the study investigated the phenomenon of exchange bias in $Mn_3Sn$/Py bilayers, revealing the potential to achieve exchange bias fields of up to $\mu_0H_{EB} = 12.6$ mT at 5 K. These results highlight the attractiveness of $Mn_3Sn$ films for applications in antiferromagnetic spintronics.

A study in [79] reports on the unusual behavior of the anomalous Hall effect in the compound $Mn_3Sn$. The authors discovered a linear dependence on the magnetic field, which they attributed to the Berry curvature of the wave function. Interestingly, the magnitude of the Hall signal in this case exceeds what would be expected in the semiclassical model. The authors propose that the magnetic field induces a nonplanar spin canting, resulting in a nontrivial chirality of the spins on the kagome lattice. This leads to changes in the band structure, specifically gapping out previously unknown Weyl nodal lines, which explains the observed behavior of the anomalous Hall effect. The findings suggest a connection between the Berry phase in real space, arising from spin chirality, and the curvature of the Berry momentum space in the kagome lattice.

The paper [80] investigates the electronic transport and magnetic properties of $Mn_3Sn$ films. The findings reveal that these films exhibit a weak uncompensated magnetic moment of approximately 0.12 $\mu_B$/f.u. and an electrical resistance of about 3.8 $\mu\Omega$ m at room temperature. These results closely match the data obtained for bulk $Mn_3Sn$, indicating high purity and perfection in the synthesized films, with a resistance ratio RRR of 3.92. The Mn atoms in the $Mn_3Sn$ compound are arranged in a kagome-type lattice. The study demonstrates that at room temperature when the external magnetic field is perpendicular to the kagome planes, a weak anomalous Hall effect is observed along with a Hall resistance that varies linearly with the field. The researchers identify three distinct magnetic phases in the investigated films of the chiral antiferromagnet $Mn_3Sn$: an inverse triangular spin state occurring above 250 K, a helical phase stabilized between 100 and 250 K, and a spin glass phase formed below 100 K. Based on their findings, the authors suggest that these films may support topologically protected skyrmions, where the fictitious effective magnetic field is estimated to be around 4.4 T. This indicates the potential for unique magnetic phenomena in the $Mn_3Sn$ films.

Currently, there is active research on the physics of Berry curvature and related nontrivial magnetic transport in two-dimensional spin-lattice kagome structures of noncollinear antiferromagnets. The focus has mainly been on hexagonal chiral antiferromagnets like $Mn_3Sn$, while studies on face-centered cubic (fcc) noncollinear antiferromagnetism are limited. One such example of fcc noncollinear antiferromagnets is $Mn_3Pt$.

In reference [81], the researchers examined single crystals of $Mn_3Pt$. They discovered that applying uniaxial stress to $Mn_3Pt$ crystals decreased their coercive force. This allowed them to observe the anomalous Hall effect in bulk materials of the $Mn_3Z$ family with a cubic structure. Interestingly, the anomalous Hall effect remained even after the stress was removed, suggesting that stress-induced ferromagnetic moments had minimal impact on the effect. The study found that longitudinal stress reduced the coercive force more than transverse stress, indicating that the dominant response to applied stress was a rotation of spins within the plane. However, further investigation is needed to verify this assumption, particularly through direct observation of piezomagnetism.

In reference [82], the authors systematically studied the effect of deformation on the magnetic properties and the anomalous Hall effect of fcc noncollinear antiferromagnetic $Mn_3Pt$ films. They found that the ferromagnetic characteristics of $Mn_3Pt$ films showed a similar response to strain, in contrast to hexagonal chiral antiferromagnets. The ferromagnetic signal in these films was attributed to the magnetic moment slope of Mn atoms in kagome structures, while the anomalous Hall resistivity was linked to the nonzero Berry curvature. By adjusting the thickness of the $Mn_3Pt$ film, they achieved an anomalous Hall conductivity exceeding 100 ($\Omega$ cm)$^{-1}$. Additionally, the relationship between the Berry phase, magnetic properties, and AHE in $Mn_3Pt$ was confirmed through a study on crystal growth orientation. The researchers concluded that this compound holds promise as a candidate for room-temperature antiferromagnetic spintronics.

**4. Physical Mechanisms of the Magnetic States Formation in Mn-Based Compounds: Half-Metallic Ferromagnets and Spin Gapless Semiconductors**

The existence of distinct spin-up and spin-down states in half-metallic ferromagnets presents a complex challenge to the general theory of itinerant magnetism [34]. The process of achieving a half-metallic state in Heusler alloys like $X$Mn$Z$ and $X_2$Mn$Z$, which have $C1_b$ and $L2_1$ structures, can be explained as follows [33,34,83,84]. When we disregard the hybridization of atomic states $X$ and $Z$, the $d$ band of manganese displays a significant energy gap between bonding and antibonding states. In a ferromagnetic state, the strong intra-atomic exchange (Hund's exchange) among manganese ions results in a notable separation of spin subbands for up and down spins. One of these spin subbands closely interacts with the ligand's $p$ band, leading to a partial or complete blurring of the corresponding energy gap due to $p$-$d$ hybridization. Meanwhile, in the other spin subband, the energy gap remains intact and has the potential to align with the Fermi level, thus giving rise to the HMF state.

The $C1_b$ structure exhibits a real energy gap, whereas the $L2_1$ structure has a pronounced pseudogap. This difference is attributed to significant alterations in $p$-$d$ hybridization, particularly between the $p$ and $t_{2g}$ states, in the absence of an inversion center, which is a characteristic feature of the $C1_b$ structure. As a result, the $C1_b$ structure is more favorable for the formation of the HMF state. The stability of the ferromagnetic state is a result of variations in $p$-$d$ hybridization between states with opposite spin orientations, as elaborated in [84].

In a study [85], researchers conducted calculations for 54 ternary Heusler compounds with the composition $X_2YZ$. In this context, $X$ represents a 3$d$ transition metal (specifically Mn, Fe, Co), while $Y$ and $Z$ represent elements Y, Zr, Nb, Mo, Tc, Ru, Rh, Pd, Ag, and Al and Si, respectively. The findings from this study revealed that seven of these compounds–namely, $Mn_2NbAl$, $Mn_2ZrSi$, $Mn_2RhSi$, $Co_2ZrAl$, $Co_2NbAl$, $Co_2YSi$, and $Co_2ZrSi$–displayed 100% spin polarization, which classifies them as half-metallic ferromagnets (HMFs). Furthermore, five other alloys, specifically $Mn_2TcAl$, $Mn_2RuAl$, $Mn_2NbSi$, $Mn_2RuSi$, and $Fe_2NbSi$, exhibited high spin polarization levels (above 90%) with a gap present for one of the spin directions near the Fermi level. These alloys were categorized as "almost HMF" in the study [85]. Importantly, their Fermi levels were found to shift under pressure, resulting in the alignment of the Fermi level with the gap and the emergence of the HMF state.

An intriguing development in the field of half-metallic magnetism is represented by electron-deficient full Heusler alloys. Reducing the number of valence electrons to 24 per formula unit leads to either a nonmagnetic semiconductor or a half-metallic antiferromagnet. Remarkably, the reduction in the number of valence electrons can continue, entering a range of half-metals but with a bandgap for the majority spin direction. This is best exemplified by the case of $Mn_2VAl$, which is a half-metallic ferrimagnet, as calculated using the generalized gradient exchange-correlation potential [86].

In addition to the mentioned study, other works (referenced as [35,86,87]) explored the exchange interactions in $Ni_2MnZ$ ($Z$ = Ga, In, Sn, Sb) and $Mn_2VZ$ ($Z$ = Al, Ge) alloys. $Ni_2MnZ$ was found to be non-half-metallic, while $Mn_2VZ$ was identified as half-metallic. These studies emphasized the significance of intersublattice exchange interactions. In the case of $Mn_2VZ$ ($Z$ = Al, Ge), it was observed that the ferrimagnetic coupling between V and Mn moments stabilized the ferromagnetic alignment of Mn moments. The V-based Heusler alloys $Mn_2VZ$ ($Z$ = Al, Ga, In, Si, Ge, Sn) were predicted to exhibit half-metallic ferrimagnetism in these studies [35,86,88].

Half-metallicity in the Mn$_2$VAl ferrimagnet was detected using resonant inelastic soft X-ray scattering (SX-RIXS) in the presence of a magnetic field [89]. When V $L$-edge excitation was applied, the findings revealed that the partial density of states for the V $3d$ states around the Fermi energy was minimal, and these states exhibited a relatively localized character. Conversely, when Mn $L$-edge excitation was used, the spectra were dominated by fluorescence and displayed clear magnetic circular dichroism, which showed significant dependence on the excitation photon energy. These experimental results were compared with theoretical predictions based on density-functional-theory band structure calculations, confirming the itinerant, spin-dependent nature of the Mn $3d$ states and the decay of the Mn $2p$ core states. This consistency in the experimental data aligns with the half-metallic behavior of the Mn $3d$ states.

The electronic structure and magnetic state of SGS materials are quite similar to HMF compounds in many ways. "Transition" from the HMF to the SGS state can occur during doping. Using the first-principles calculations within density functional theory, the authors of ref. [90] described such a transition by studying the magnetism and electronic structure of Co$_{2-x}$Mn$_{1+x}$Al ($0 \leq x \leq 1$) alloys, i.e., during the transition from HMF Co$_2$MnAl to SGS Mn$_2$CoAl.

As reported in [90], the electronic spectrum of the Co$_2$MnAl compound ($x = 0$) exhibits an intersection of its spin-up and spin-down bands with the Fermi level $E_F$, indicating metallic properties for both spin directions. However, Co$_2$MnAl displays a low density of states for spin-down electrons, resulting in a pseudogap rather than a true gap. The calculated spectra show a spin polarization of approximately 75%. As the manganese content increases, the spin-down valence band of Co$_2$MnAl shifts downward, causing the density of states to approach zero. Consequently, the degree of spin polarization increases. In the case of the Co$_{1.5}$Mn$_{1.5}$Al compound, as Mn content further increases, the top of the spin-down band aligns with the Fermi level, leading to the emergence of a genuine gap and achieving a maximum spin polarization of 100%. Despite that, Co$_{1.5}$Mn$_{1.5}$Al faces a vulnerability as its Fermi level is situated at the edge of the spin-down bandgap, making it susceptible to external influences, especially structural defects.

As Mn content continues to rise in Co$_{2-x}$Mn$_{1+x}$Al, the spin-down valence band moves further away from the Fermi level, resulting in an enlarged band gap. At $x = 1.875$, the width of the spin-down gap reaches approximately 0.4 eV, with the Fermi level nearly positioned at the midpoint of the gap. This leads the authors of reference [90] to classify the Co$_{1.125}$Mn$_{1.875}$Al compound as an "ideal" half-metallic ferromagnet. With increasing Mn content to $x = 2$, the compound transforms into an inverse Heusler alloy Mn$_2$CoAl. In this compound, a genuine energy gap is preserved for spin-down states. The electronic structure of Mn$_2$CoAl exhibits a unique characteristic: for spin-up electronic states, the conduction band and valence band edges closely approach the Fermi level, resulting in a minimal energy gap. This distinctive feature categorizes the inverse Mn$_2$CoAl compound as a spin gapless semiconductor.

Table 1 presents information on Mn$_2$YZ and Mn$_3$Z (Y = Cr, Fe, Co; Z = Al, Ge, Si, Sn, Pt), for which the electronic state, magnetic state, and anomalous Hall conductivity are given.

**Table 1.** Manganese-based compounds of Mn$_2$YZ and Mn$_3$Z (Y = V, Cr, Fe, Co, Ni; Z = Al, Ge, Si, Sn, Pt, Ir). Electronic and magnetic states and anomalous Hall conductivity.

| Compound | Electronic State | Magnetic State | Anomalous Hall Conductivity $\sigma^{AHE}$ (Ohm cm)$^{-1}$ |
|---|---|---|---|
| Mn$_2$VAl | HMF [1] [41] | FIM [2] [41] | |
| Mn$_2$VSi | HMF [41] | FIM [41] | |
| Mn$_2$CrAl | HMF [91,92] | CFIM [3] [21] | |
| Mn$_2$FeAl | HMF [36] | CFIM [42]; Frustrated AFM [4] [44] | |
| Mn$_2$NiAl | | AFM [93] | |
| Mn$_2$CoAl | SGS [5] [47] | FM [6] [47] | 21.8 [47] |
| Mn$_3$Al | HM-AFM [7] [94] | CFIM (cast); frustrated AFM (RMQ) [8] [28,29]; AFM [94] | 320 [62] |
| Mn$_3$Ge | HM-FIM [9] [94]; TSM [10] [9] | FIM [94]; Noncollinear AFM [9,63] | 60 at RT; ~380 at 5 K [63] |
| Mn$_3$Si | HM-FIM [94] | FIM [94] | |
| Mn$_3$Sn | TSM [79] | Noncollinear AFM [78,81]; kagome AFM [79] | ~20 at RT; ~100 at 100 K [95] |
| Mn$_3$Pt | TSM [82] | kagome AFM [82] | 100 [82] |
| Mn$_3$Ir | | AFM [96] | 40 at RT [96] |

[1] HMF is a half metallic ferromagnet. [2] FIM is a ferrimagnet. [3] CFIM is a compensated ferrimagnet. [4] AFM is an antiferromagnet. [5] SGS is a spin gapless semiconductor. [6] FM is a ferromagnet. [7] HM-AFM is a half-metallic antiferromagnet. [8] RMQ is rapid melt quenched. [9] HM-FIM is a half-metallic ferrimagnet. [10] TSM is a topological semimetal.

## 5. Conclusions

Intermetallic compounds based on manganese, specifically Mn$_2$YZ and Mn$_3$Z (where (Y = Sc, Ti, V, Cr, Fe, Co, etc.; Z = Al, Ge, Sn, Si, Pt, etc.), exhibit unique characteristics of the electronic structure. These compounds have the potential to exhibit various electronic and magnetic states, such as antiferromagnet, compensated ferrimagnet, topological semimetal, and frustrated antiferromagnet. Furthermore, their magnetic and electronic properties are highly sensitive to external influences.

In the electronic subsystem of Mn$_2$YAl compounds, different magnetic states can lead to the emergence of half-metallic ferromagnet and spin gapless semiconductor states. For instance,

Mn$_2$CoAl can exhibit a ferromagnetic spin gapless semiconductor state, while Mn$_2$FeAl can display a frustrated antiferromagnet state.

Compounds of Mn$_3$Z manifest as a half-metallic ferromagnet or a topological semimetal, displaying notable phenomena such as a large anomalous Hall effect, spin Hall effect, spin Nernst effect, and thermal Hall effect. The magnetic subsystem of these compounds can exhibit states such as ferrimagnetism, antiferromagnetism, compensated ferrimagnetism, frustrated antiferromagnetism, and antiferromagnetism with a kagome-type lattice. Overall, Mn$_3$Z compounds offer a rich spectrum of electronic and magnetic properties.

When comparing the properties of Mn compounds Mn$_2$YAl and Mn$_3$Z, it becomes evident that there are both similarities and differences between them, as shown in Table 1. In the electronic structure of Mn$_2$YAl alloys, the predominant state is that of a half-metallic ferromagnet, whereas in the Mn$_2$CoAl compound, it exhibits characteristics of a spin gapless semiconductor. It is worth noting that Mn$_2$CoAl is among the first Heusler alloys where the SGS state is prominently observed. In terms of magnetism, Mn$_2$YAl alloys display a range of behaviors, including ferromagnetism, compensated ferrimagnetism, and frustrated AFM state.

On the other hand, the Mn$_3$Z compounds exhibit more diverse electronic and magnetic states. Many of them demonstrate the characteristics of a topological semimetal with significant anomalous Hall conductivity, even at room temperature. Similar to Mn$_2$YAl alloys, they also exhibit a variety of magnetic states. Moreover, due to their richer electronic structure, they can realize states such as noncollinear AFM states and AFM states typical for a kagome-type lattice.

It should be emphasized that in actual materials, the presence and behavior of electronic and magnetic states, as well as transitions between them, can vary depending on various factors. These factors include the composition of the sample, external parameters such as temperature, magnetic field, and external pressure, as well as the dimensionality of the material (bulk, film, nanostructure), and the processing method used (cast, rapidly quenched, nanostructured samples, etc.). The electronic and magnetic states discussed, as well as the transitions between them, can be implemented and put into practice.

Besides the practical applications, all the states discussed, especially frustrated antiferromagnetism and half-metallic ferromagnetism, are of general scientific interest. Some systems considered can demonstrate anomalous electronic properties (particularly, large $T$-linear heat capacity) and are highly intriguing from a physical standpoint, posing an exciting challenge for future theoretical advancements.

It should be noted that a significant number of studies on manganese-based Heusler compounds exploit band calculations, which may not always be reliable. This is especially true when treating the size of the energy gap and the degree of spin polarization in the case of possible HMF and SGS compounds, thereby overestimating the stability of these states. However, there are ongoing improvements in electronic structure calculation methods, and new modern techniques are emerging. It is essential to constantly compare these calculations with experimental data. For example, in the case of the HMF, this could involve comparing the results of band calculations and direct experiments to determine the gap and spin polarization in situ by ultraviolet-photoemission spectroscopy, taking advantage of a multichannel spin filter [97]. NMR experiments could also be conducted to search for the vanishing of a linear contribution to the temperature dependence of the nuclear spin-lattice relaxation rate $1/T_1$ (violation of Korringa's law), which is theoretically predicted for the HMF [34].

Finally, a promising direction for exploring new electronic effects and magnetic state features is a comprehensive study of quaternary Heusler alloys containing manganese atoms. Although the present review did not cover such compounds, there are many studies in this area, which can be found in related reviews [31,98,99] and references therein.

This opens up further opportunities for fine control of the magnetic and electronic characteristics of such compounds for their possible practical use in spintronics and micro- and nanoelectronics.

**Author Contributions:** Conceptualization, V.V.M. and V.Y.I.; methodology, V.V.M. and V.Y.I.; formal analysis, V.V.M. and V.Y.I.; investigation, V.V.M.; resources, V.V.M. and V.Y.I.; data curation, V.V.M. and V.Y.I.; writing—original draft preparation, V.V.M. and V.Y.I.; writing—review and editing, V.V.M. and V.Y.I.; visualization, V.V.M. and V.Y.I.; supervision, V.V.M. and V.Y.I.; funding acquisition, V.V.M. All authors have read and agreed to the published version of the manuscript.

**Funding:** The research was carried out within the state assignment of Ministry of Science and Higher Education of the Russian Federation (themes «Spin», № 122021000036-3 and «Quantum», № 122021000038-7). Section 3 "Features of the electronic transport and magnetic state of Mn$_3$Z (Z = Al, Ge, Si, Sn, Pt etc.) compounds" was prepared with the financial support of the Russian Science Foundation within the framework of research project No. 22-22-00935.

**Acknowledgments:** The authors consider it their pleasant duty to thank their colleagues and co-authors Lukoyanov, A.V., Skryabin, Y.N., Marchenkova, E.B. for valuable discussions, and Perevozchikova, Y.A., Perevalova, A.N., Fominykh, B.M., Semiannikova, A.A., Emelyanova, S.M. for help with the design of the review.

**Conflicts of Interest:** The authors declare no conflict of interest.